\documentclass[ amsmath, amssymb, breqn, aps, prfluids, superscriptaddress, onecolumn, longbibliography]{revtex4-2}
\usepackage{graphicx}
\usepackage{amsmath,amssymb}
\usepackage{textcomp}
\usepackage{hyperref}
\usepackage[dvipsnames]{xcolor}
\usepackage{soul}

\newcommand{\bnabla}{{\mbox{\boldmath $\nabla$}}}
\newcommand{\bom}{{\mbox{\boldmath $\omega$}}}

\newcommand{\vLc}{{v}_L^\text{crit}}
\newcommand{\vrc}{{v}_\text{rot}^\text{crit}}
\newcommand{\vvc}{{v}_v^\text{crit}}
\newcommand{\vc}{{v}^\text{crit}}
\newcommand{\Mc}{{M}^\text{crit}}
\newcommand{\MLc}{{M}_L^\text{crit}}
\newcommand{\Mrc}{{M}_\text{rot}^\text{crit}}
\newcommand{\Mvc}{{M}_v^\text{crit}}
\newcommand{\dvc}{\Delta{v}^\text{crit}}
\newcommand{\xx}{{\bf {x}}}
\newcommand{\kk}{{\bf {k}}}
\newcommand{\vv}{{\bf {v}}}

\newcommand{\angstrom}{\mbox{\normalfont\AA}}
\newcommand{\Vext}{V_\text{ext}}
\newcommand{\krot}{k_\text{rot}}
\newcommand{\Uvec}{\mathbf{U}}

\def\etal{\textit{et al. }}

\usepackage{soul}
\def\3He{$^3$He}
\def\4He{$^4$He}
\begin{document}

\title{Pressure and Size Dependence of Roton Emission and Vortex Creation by Moving Objects in He~II in $T \to 0$ Limit:  Generalized Nonlocal Gross-Pitaevskii Model }

\author{Nicol\'as P. M\"uller}
\affiliation{Laboratoire de Physique des Plasmas (LPP), CNRS, Observatoire de Paris, Sorbonne Université, Université Paris-Saclay, École polytechnique, Institut Polytechnique de Paris, 91120 Palaiseau, France}
\email{nicolas.muller@lpp.polytechnique.fr}

\author{Ladislav Skrbek}
\affiliation{Faculty of Mathematics and Physics, Charles University, Ke Karlovu 3, Prague, 121 16, Czech Republic}
\email{ladislav.skrbek@matfyz.cuni.cz}

\author{Yuri A. Sergeev}
\affiliation{School of Mathematics, Statistics and Physics, Newcastle University, Newcastle upon Tyne NE1 7RU, United Kingdom}
\email{yuri.sergeev@newcastle.ac.uk}

\author{Giorgio Krstulovic}
\affiliation{Universit\'{e} C\^{o}te d'Azur, CNRS, Institut de Physique de Nice (INPHYNI), 17 rue Julien Lauprêtre, 06200 Nice, France}
\email{giorgio.krstulovic@univ-cotedazur.fr}

\date{\today}

\begin{abstract}
In the framework of generalized, nonlocal Gross-Pitaevskii (GP) model, we study numerically the pressure- and size-dependent mechanisms of roton emission and vortex nucleation by objects moving in superfluid \4He. As far as the authors are aware, this is the first attempt to analyze the pressure dependence of these mechanisms and the associated critical velocities within a single theoretical framework. For each of several pressures in the range from 0 to the solidification pressure of $\approx25$~bar, we chose the parameters of the interatomic interaction potential such that the resulting excitation spectrum for the generalized, nonlocal GP equation approximates fairly accurately the pressure-dependent dispersion curve determined experimentally by Godfrin \textit{et al.}, Phys. Rev. B \textbf{103}, 104516 (2021). In the two-dimensional approximation, for circular obstacles (disks) moving in quiescent \4He, we calculated two critical velocities---one corresponding to the roton emission and the other to the nucleation of quantized vortices---as functions of pressure and the obstacle's size.{We also comment briefly on three-dimensional simulations of the roton emission and vortex nucleation by moving spherical obstacles.} 
\end{abstract}
\maketitle

\section*{preface}
\label{sec:preface}

Cryogenic phase of $^4$He, the common isotope of helium, belongs to the most investigated fluids in nature. Its liquefaction by Kamerlingh-Onnes on July 10, 1908, marks the beginning of modern low temperature physics. The discovery of superfluidity by Kapitza \cite{Kapitza} and Allen \cite{Allen} opened the path to investigations of the new class of condensed matter systems, the quantum fluids, whose properties cannot be described by Navier-Stokes equations but depend on quantum mechanics. The extraordinary properties of the superfluid phase of \4He, for historical reasons known as He~II, are at present fairly well understood, thanks to influential works of early investigators. They include theoretical work of London \cite{London} who linked superfluidity with Bose-Einstein condensation, introduction of the two-fluid model (experimentally verified by Andronikashvili \cite{Andronikashvili}) and prediction of second sound by Tisza \cite{Tisza} and Landau \cite{Landau41,Landau47}; second sound then experimentally discovered by Peshkov \cite{Peshkov1,Peshkov2}. Onsager \cite{Onsager} predicted the existence of line singularities---quantum vortices in the superfluid component of He~II, whose  circulation is (usually singly \cite{VinenSingle}) quantized in units of $\kappa=2\pi \hbar /m_4$, where $\hbar$ is Planck's constant and $m_4$ is the mass of the superfluid particle, the $^4$He atom. Their existence leads to quantum turbulence, first considered as a theoretical possibility by Feynman \cite{Feynman} and investigated in pioneering experiments in thermal counterflow of He II by Vinen \cite{VinenOld}. Landau \cite{Landau41,Landau47} predicted the shape of the dispersion curve as consisting of linear ``phonon'' part at small $k$ followed by a broad local maximum and  ``roton'' minimum. Cohen and Feynman \cite{FeynmanCohen56} suggested to determine the dispersion curve experimentally by using inelastic neutron scattering. A number of experiments confirming the Landau's prediction followed \cite{Glyde2018,GodfrinDispersion}.

A moving body, such as a nearly macroscopic positive ion ``snowball'' or negative ion ``bubble'', accelerated in He~II \cite{Meyer61} by an electric field to a certain critical velocity, can generate excitations---rotons, phonons and/or quantized vortices \cite{RR64,Rayfield66} ({or, alternatively, cause cavitation \cite{SkrbekSergeevCritVel})---destroying superfluidity. Fundamental critical velocities for roton generation and vortex creation in isotopically pure bulk He~II in the zero temperature limit have been investigated by McClintock and coworkers in a series of experiments in Lancaster \cite{Phillips73,Allum77,Stamp79,Nancolas82,Ellis82,Bowley82,Ellis85,Nancolas85-PhilTrans,Nancolas85,Hendry88,Hendry90} (see also a detailed review by McClintock and Bowley \cite{McClintockBowleyProgressLT95}) and theoretically by Bowley and Sheard \cite{Bowley77}, Bowley \cite{Bowley84}, and Muirhead, Vinen and Donnelly \cite{Muirhead} (see also Ref.~\cite{BarenghiMcClintockMuirhead}). We should also mention here the fundamental theory of roton creation by moving ions developed earlier by Iordanskii \cite{Iordanskii68} and Volovik \cite{Volovik70}. At low externally applied pressures, negative ions created charged vortex rings, while at high pressures above 10~--~11~bar the drag on a moving negative ion remains negligibly small until a critical velocity, $\vrc$ is attained, which is close to the Landau critical velocity
\begin{equation}
\vLc\approx\Delta_\text{R}/\hbar k_0\,,
\label{eq:vL}
\end{equation}
where  $\Delta_\text{R}$ and $k_0$ are, respectively, the pressure dependent roton energy gap and the wave vector $k_0$ at the roton minimum~\cite{Brooks77,GodfrinDispersion}, and then quickly rises, until at another critical velocity vortex generation occurs. These Lancaster time of flight experiments are considered as low temperature classics. A pressure dependence of the ion's critical velocity $\vrc$ corresponding to the emergence of roton emission was correctly attributed by McClintock, Bowley and their co-workers (see e.g. Ref.~\cite{Bowley77} and the review by McClintock and Bowley \cite{McClintockBowleyProgressLT95} and references therein) to the pressure dependence of the Landau critical velocity given by Eq.~(\ref{eq:vL}). However, $\vrc$  does not exactly coincide with $\vLc$, see e.g. Ref. \cite{Ellis85} for detail. 

The pressure dependence of the critical velocity of vortex creation, $\vvc$ was known from the results of experimental measurements~\cite{Rayfield68,Stamp79,Bowley82,Hendry90}, backed up by semi-classical theoretical calculations of Muirhead \textit{et al.} \cite{Muirhead} (see also Ref.~\cite{BarenghiMcClintockMuirhead}) and Bowley \cite{Bowley84}. Note, however, that the values of $\vvc$ predicted by Bowley's calculation were higher than those calculated by Muirhead \textit{et al.} and observed experimentally.

The study presented in this article is motivated by two recent developments, one theoretical and the other experimental. The former is an application of the generalized, nonlocal Gross-Pitaevskii (GP) model to the study of roton emission and nucleation of quantized vortices by objects moving in superfluid \4He \cite{Muller20,Muller22}. The latter is a detailed experimental study by Godfrin \etal\cite{GodfrinDispersion} of the properties of superfluid \4He at elevated pressures; this study resulted, among other important findings, in the accurate quantitative description of the Landau dispersion curve for pressures ranging from $P=0$ to $P=24$~bar, the latter being close to the pressure of solidification at $\approx25$~bar. Our article, after discussing the situation in the historical context, offers a physically motivated approach which enables us to analyze the physical mechanisms responsible for the pressure dependence of two critical velocities, $\vrc$ and $\vvc$ within a single theoretical framework based on the generalized, nonlocal GP model. Because a numerical analysis of this model presents substantial difficulties, our study is mainly restricted by the analysis of the modeling, two-dimensional (2D) problem (although we comment, rather briefly, on three-dimensional (3D) calculations as well). Nevertheless, the results of our 2D numerical simulations explain the underlying physics of the problem in qualitative agreement with experimental observations. 

\section{introduction}
\label{sec:intro}

\subsection{Dispersion relation at different pressures}
\label{sec:dispersion}

The most fundamental property of a many-body system is represented by the dispersion relation for elementary excitations, $E(k)$, where $E$ and $k$ are their energy and wavenumber, respectively. For superfluid $^4$He at saturated vapor pressure (SVP), \citet{Landau41,Landau47} predicted the shape of the dispersion curve as consisting of linear ``phonon'' part at small $k$ followed by a broad local maximum at $k\approx1\,\angstrom^{-1}$ (``maxon'') which, most importantly, is followed in turn by a deep ``roton'' minimum at $k\approx2\,\angstrom^{-1}$. 

Following the original proposition by Landau, Feynman linked the emergence of rotons to the transition from a vortex to an elementary excitation. To quote Ref.~\cite{Feynman}, ``a roton is the ghost of a vanishing vortex ring''. The nature of rotons continues to be discussed, see e.g. Refs. \cite{Nozieres2004,Balibar2007}. In the latter of these works, Balibar showed that a sharp maximum in the static structure factor of liquid helium at the wavenumber $k$ corresponding to the average interatomic distance is directly linked to the roton minimum of the dispersion curve, and hence concluded that rotons, whose existence works against superfluid order, are precursors to solidification. 

Later elaborate experiments  (see e.g. Ref.~\cite{GodfrinDispersion} and references therein) confirmed the shape of dispersion curve suggested by Landau and, importantly, resulted in the detailed description of $E(k)$ and roton and maxon parameters in dependence of temperature and pressure, the latter in the range from SVP to $P=24.08$~bar, close to the pressure of solidification at $\approx25$~bar. At $P=0$ and temperature $T<0.1$~K, the dispersion curve obtained by neutron scattering~\cite{GodfrinDispersion} is shown in Fig.~\ref{fig:Godfrin}.
\begin{figure}[h]
\centerline{\includegraphics[width=.59\linewidth]{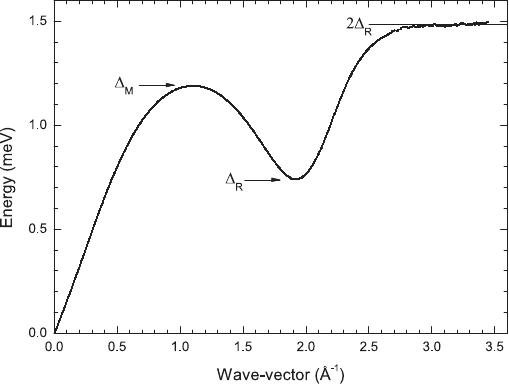}}
\medskip
\caption{The dispersion curve of $^4$He at $P=0$ and $T<0.1$~K determined by neutron scattering. $\Delta_\text{M}$ and $\Delta_\text{R}$ are the maxon and roton energies (``gaps''). The plateau at $E=2\Delta_\text{R}$ ($k\gtrsim2.8~\angstrom^{-1}$) is known as the ``Pitaevskii plateau'', see e.g. a review~\cite{Glyde2018}. Reprinted by permission from Godfrin \textit{et al.}, Phys. Rev. B \textbf{103}, 104516 (2021). \textcopyright2021 American Physical Society.}
\label{fig:Godfrin}
\end{figure}
The dispersion relation at high wave vectors increases up to a maximum energy $\approx 2 \Delta_\text{R}$, called the Pitaevskii plateau \cite{PiaevskiiPlateau}. Since high-$k$ excitations can decay into roton pairs, they cannot remain sharp above the plateau. The experimental investigation of this region is difficult (see~\cite{GodfrinDispersion} and references therein). At saturated vapor pressure, single excitations reach twice the roton gap at 2.8 inverse \AA, the energy remains constant in the vicinity of $2 \Delta _\text{R}$ between 3.0~\AA$^{-1}$, and the end point of the dispersion curve is at $k = 3.6$~\AA$^{-1}$.
As for even higher $k$, Marchenko and Parshin \cite{MarchenkoParshin} claimed that the spectrum of the bulk excitations should be
recovered at a certain critical point with the coordinates of about several roton energies and momenta in the
form of the spectrum of vortex rings. 

Landau's critical velocity provides a general expression for the breakdown of superfluidity. In particular, it states that when an object moves at a velocity ${\bf v}$, quasiparticles are excited at wavevectors ${\bf k}$ such that
\begin{equation}
   \hbar{\bf v}\cdot{\bf k}+E({\bf k})=0\label{eq:LandauCriterium}. 
\end{equation}
In the case of superfluid helium, it follows from Fig.~\ref{fig:Godfrin} that this equation has a solution only if the velocity of the object exceeds the one given by the roton minimum determined by formula \eqref{eq:vL}. 
For this work, it is thus important to know how the roton minimum shifts with externally applied pressure, which is shown in detail in Fig.~\ref{fig:RotonMin}. We emphasize that these experimental data are backed up by calculations in the framework of the density functional theory \cite{GodfrinDispersion}. 
\begin{figure}[h]
\centerline{\includegraphics[width=.59\linewidth]{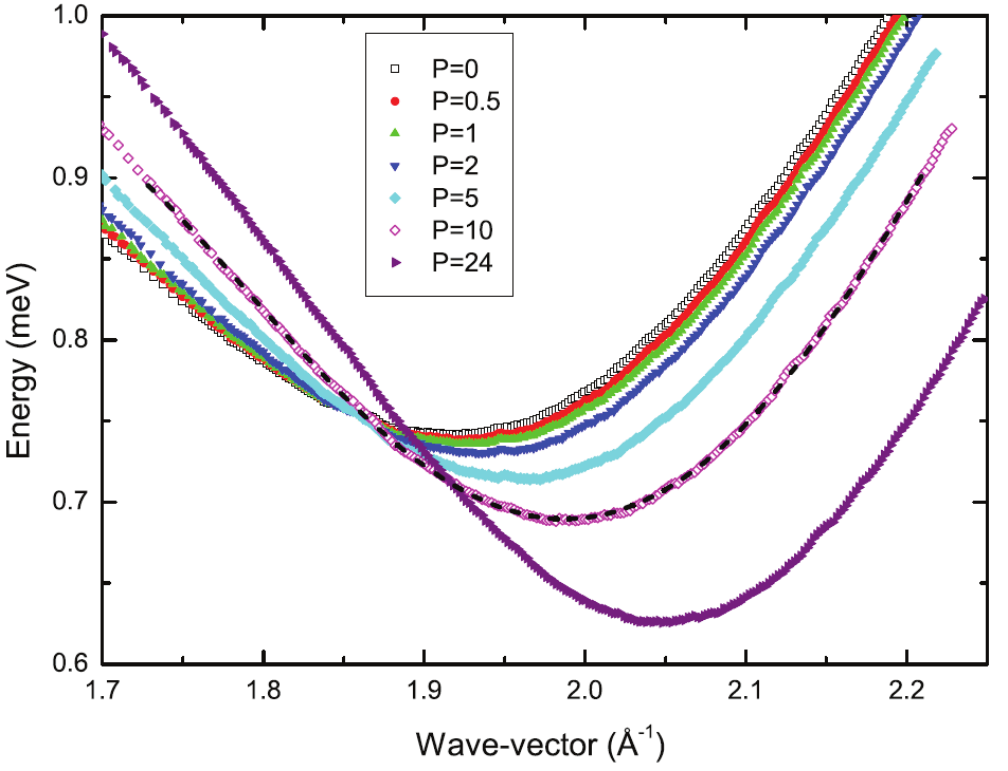}}
\medskip
\caption{The dispersion relation in the vicinity of the roton minima
at different pressures (accurate values of pressure are: 0; 0.51; 1.02;  2.01; 5.01; 10.01 and 24.08 bar). 
Reprinted by permission from Godfrin \textit{et al.}, Phys. Rev. B \textbf{103}, 104516 (2021). \textcopyright2021 American Physical Society.}
\label{fig:RotonMin}
\end{figure}

\subsection{Pressure dependence of critical velocities. A short review of earlier experimental and theoretical studies}
\label{sec:critvelocities}

Experimental verification of the Landau critical velocity $\vLc$ as well as the  critical velocity for vortex nucleation $\vvc$ required to overcome a number of hurdles. The critical velocities observed in numerous early experiments were much lower than those theoretically predicted, sometimes by orders of magnitude.  There are several important issues that must have been taken into account.

Firstly, almost all macroscopic samples of He~II contain a few vortices with their ends pinned on the container walls or on solid objects within the superfluid; \citet{Awschalom84} even provided an estimate of their density. The remnant vortices might have been left from an earlier experiment, or they might have been formed by the Kibble–Zurek mechanism \citep{Zurek} when the helium sample was cooled through the $\lambda$-transition. It is these, so called remnant vortices, that in practice are usually involved in the breakdown of superfluidity \cite{Vinen14} upon exceeding the flow velocities typically of order few cm/s.
There are several ways to avoid the influence of remnant vortices. One can try to carefully prepare a macroscopic vortex free sample of He~II, by slowly condensing $^4$He through a tiny pinhole \cite{Goto} or to probe a very small volume, likely containing no remnant vortices. To this end, efficient probes are helium ions,  so small that their motion is unlikely affected by remnant vorticity. The ions can be thought of as very small charged spheres that may be manipulated with electric fields and detected by the currents they induce in surrounding electrodes. The positive ion of radius $\approx 0.6\,$nm  can be described as a snowball of about 20 helium atoms surrounding a $^4$He$^+$ ion, held together by electrostriction. The negative ion is an electron within an otherwise empty bubble. Its radius, $R$, can be estimated by minimizing the energy $E$ of this object, assuming that this energy consists of the zero point energy, surface energy of the bubble  and the work done against the liquid pressure in creating the cavity \cite{Maris}:
\begin{equation}
E=\frac{(2\pi\hbar)^2}{8mR^2}+4\pi R^2\alpha + \frac{4}{3}\pi R^3 P\,,
\label{eq:E}
\end{equation}
\noindent where $m$ is the electron mass and $\alpha$ denotes the surface tension of He~II. At zero temperature (i.e., in the $T \to 0$ limit), minimizing $E$ given by Eq.~(\ref{eq:E}) yields (depending on the exact value of $\alpha$) the radius $R$  slightly below 2~nm \cite{Maris}. Experimentally, Springett and Donnelly \cite{SpringettDonnelly} derived the pressure dependence of the radius of the negative ion from the measurements of ion's trapping cross section on quantized vortex lines. Ostermeier \cite{Ostermeier} performed accurate measurements of the mobility of positive and negative charge carriers in He~II, at pressures up to the melting pressure and temperatures down to 0.27 K, and analyzed them using existing theories which yielded the electron-bubble radius as a function of pressure. Generally, the radius of the negative ion has been found to vary between $\approx 1.1\,$~nm (near the melting pressure) and $\approx 1.7\,$nm (under equilibrium saturated vapor pressure), see Fig.~\ref{fig:IonRad}.
\begin{figure}[h]
\centerline{\includegraphics[width=.59\linewidth]{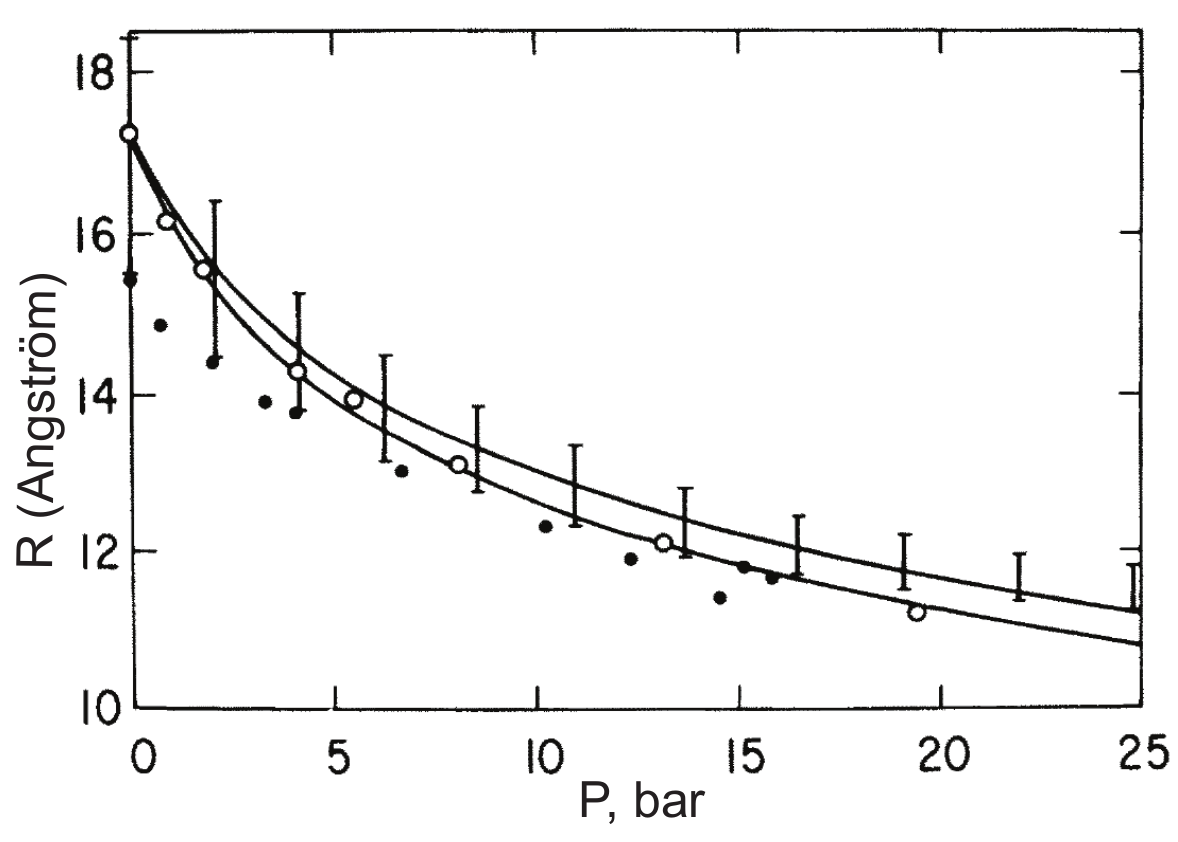}}
\medskip
\caption{Variation of the negative ion radius with pressure. Shown are the data derived from measurements of trapping cross section in quantized vortex lines by Springett and Donnelly as well as the data derived from the measurements of the mobility of negative charge carriers in He~II. For details, see  Ostermeier  Phys. Rev. A \textbf{8} 514–529 (1973) (Ref.~\cite{Ostermeier}), where the original figure is reprinted from by permission.
\textcopyright1973  American Physical Society.}
\label{fig:IonRad}
\end{figure}

Secondly, a crucial factor for reliable experimental investigation of critical velocities is the purity of the He~II sample. Impurities other than $^3$He can be removed  relatively easily via freezing them out, but at temperatures well below $1\,$K even tiny concentration of $^3$He atoms lowers the observed critical velocities significantly \citep{Nancolas85}. Concentration of $^3$He  atoms in technical $^4$He, typically around 1 ppm, is too high for this kind of sensitive experiments performed at very low temperature. To this end, McClintock's group in Lancaster \citep{McClintock78,Hendry87} developed a purification cryostat capable of producing, in continuous operation, $^4$He of excellent isotopic purity, probably better than $5 \times 10^{-13}$, suitable for such experiments. This purified $^4$He has then been supplied to other laboratories worldwide; in other laboratories the investigators purified the He~II sample in situ using the heat flush effect of thermal counterflow when filling their experimental cells.

We now describe in more detail the experimental studies of pressure dependences of the roton emission and the nucleation of vortex rings by moving objects (such as negative ion bubbles) in low temperature, pure $^4$He.
The earliest experimental studies of motion, in the electric field, of negative ions in pressurized $^4$He were carried out in 1960s by Meyer and Reif \cite{Meyer61} and Rayfield \cite{Rayfield66, Rayfield68}. The latter found that the behavior of negative ions at pressures higher than $\approx12$~bar differed from that at lower pressures. In particular, Rayfield discovered that at high pressures ions could be accelerated to the critical velocity which was close to $\vLc$ and decreased with pressure. Making use of the available data \cite{Brooks77} for the pressure dependences of the roton parameters $\Delta$ and $k_0$, it can immediately be seen that such a behavior is consistent with the pressure dependence of the Landau critical velocity predicted by Eq.~(\ref{eq:vL}). However, Rayfield's experiments \cite{Rayfield66, Rayfield68} had a serious drawback: They were carried out at temperature $T=0.6$~K. At this, relatively high temperature a drag force exerted on the ion bubble by the scattering of ambient rotons is not negligible~\cite{normal}.

Experimental studies of negative ion currents and individual ion's motion in the electric field at lower helium temperatures, $0.27~\text{K}\lesssim T\lesssim0.5~\text{K}$ such that the drag force caused by scattering of ambient quasiparticle excitations can be neglected, were conducted in Lancaster by McClintock and his coworkers and reported in numerous publications \cite{Phillips73,Allum77,Stamp79,Nancolas82,Ellis82,Bowley82,Ellis85,Nancolas85-PhilTrans,Nancolas85,Hendry88,Hendry90,McClintockBowleyProgressLT95}. These experiments were performed in the range of pressures from $P=0$ to the pressures close to that of solidification at $P\approx25$~bar. In the earliest of these works, Phillips and McClintock \cite{Phillips73} found that at pressure $P_\text{crit}\approx10$~bar the ion current rapidly increased with pressure and, for $T<0.5$~K, became independent of temperature. Based on the interpretation of their later experiments \cite{Allum77} supported by theoretical considerations by Bowley and Sheard \cite{Bowley77}, McClintock and his coworkers concluded \cite{Allum77} that at pressures below $P_\text{crit}$ most of the ions in the current traveled as charged vortex rings \cite{charged_rings}. As pressure increased beyond $P_\text{crit}$, a larger proportion of the ion current consisted of \textit{bare ions} (more precisely, ion bubbles) traveling with the velocity $\vrc$ close to the Landau critical velocity $\vLc$. Further theoretical studies \cite{Bowley84,Muirhead} of the nucleation of vortex rings and measurements, in the wide range of external pressures, of the critical velocities for roton creation \cite{Ellis85} by ions moving in the electric field enabled McClintock, Bowley and their co-workers to arrive at the following interpretation \cite{Nancolas85,McClintockBowleyProgressLT95} of their findings: The reason why an elevated external pressure is needed to support a current of bare ions is the pressure dependencies of the Landau critical velocity $\vLc$ (more precisely, of the critical velocity of roton emission $\vrc$ which differs slightly from $\vLc$; for details of the difference between $\vrc$ and $\vLc$ see Ref. \cite{Ellis85}), and of the critical velocity of nucleation of the vortex rings, $\vvc$. At a given pressure, the ion can be accelerated by an electric field to the lowest of the two critical velocities, $\vrc$ or $\vvc$.  The Landau critical velocity $\vLc$ and the critical velocity of roton emission $\vrc$ decrease with pressure. Experimental \cite{Bowley82,Hendry90} and theoretical \cite{Bowley84,Muirhead} investigations of the critical velocity of rings' creation showed that $\vvc$ increases with pressure (note that the theory developed by Muirhead, Vinen, and Donnelly \cite{Muirhead} agrees with experimental data better than that proposed by Bowley \cite{Bowley84}). Remarkably, the theoretical curves for $\vrc(P)$ and $\vvc(P)$ cross at 10 -- 12~bar (see Fig.~1 in Ref. \cite{Ellis85}) in good agreement with the critical pressure $P_\text{crit}$ observed experimentally. These findings made clear the reason for a difference in the character of ions' motion at high and low pressures: At pressures above $P_\text{crit}$, upon reaching the velocity $\vrc(P)$ the motion of ions becomes controlled by the energy dissipation (and hence the drag force) caused by the emission of rotons. At pressures below $P_\text{crit}$, the limiting velocity $\vvc$ is associated with dissipation caused by generation, by a moving ion bubble, of a different type of excitations---quantized vortex rings 
and the experiment cannot determine $\vrc(P)$ here. On the other hand, at high pressures above $P_\text{crit}$, in strong enough electric fields, the negative ions can be forced to drift with the velocity exceeding $\vrc(P)$, which allowed the Lancaster group to estimate both $\vrc(P)$ for roton emission and higher $\vvc(P)$ for nucleation of vortex rings \cite{Bowley82, Hendry90}.

Unfortunately, the absence of a mathematically simple microscopic theory of superfluid $^4$He precludes a description of these two pressure-dependent dissipative mechanisms and critical velocities associated with them within a single theoretical framework. As an alternative, we suggest below an approach based on the generalized, nonlocal Gross-Pitaevskii model which incorporates the correct pressure-dependent dispersion relation for $^4$He and also reproduces some of the other essential properties of superfluid helium.

\section{Modeling superfluid He II by the generalized, nonlocal Gross-Pitaevskii equation}
\label{sec:gnlGPE}

Because the origin of $^4$He superfluidity is associated with the Bose-Einstein condensation of bosonic helium atoms, the superfluid can be described, in the mean-field approximation, in terms of the complex order parameter---macroscopic wave function
\begin{equation}
\psi(\xx,\,t)=\sqrt{n(\xx,\,t)}\exp[i\theta(\xx,\,t)]\,,
\label{eq:psi}
\end{equation}
where $\xx$ is the spatial coordinate, $n=\rho_0/m_4=\vert\psi\vert^2$ is the number density of the condensed atoms (in the zero-temperature limit considered in this work, $\rho_0=\rho$, where $\rho=0.145~\text{g/cm}^3$ is the density of helium), and $\theta$ is the phase whose gradient determines the superfluid velocity, $\vv=(\hbar/m_4)\bnabla\theta$. The superflow is irrotational, $\bnabla\times\vv={\bf 0}$, but superfluids can support line defects---quantum vortices on which the vorticity, $\bom=\bnabla\times\vv$ is singular. Each vortex carries a single quantum of circulation, $\kappa=2\pi\hbar/m_4\approx10^{-3}~\text{cm}^2/\text{s}$.

In what follows, a phenomenological description of superfluid \4He will be based on the generalized, nonlocal Gross-Pitaevskii equation (gnlGPE) proposed by Berloff and Roberts \cite{BerloffRobertsVI-1999} in the form
\begin{widetext}
\begin{equation}\label{eq:gnlGPE}
i\hbar\frac{\partial\psi}{\partial t}=\left[-\frac{\hbar^2}{2m_4}\nabla^2-\mu(1+\chi)+g\int V(\vert\xx-\xx'\vert)\vert\psi(\xx',\,t)\vert^2\,d\xx' 
+g\chi\frac{\vert\psi\vert^{2(1+\gamma)}}{n_0^\gamma}+V_\text{ext}(x,\,t)\right]\psi\,,
\end{equation}
\end{widetext}
where $\mu$ is the chemical potential, $g=4\pi\hbar^2a_s/m_4$ is the constant of interatomic interactions, with $a_s$ being the $s$-wave scattering length, $n_0=\vert\psi_0\vert^2$ is the density of bosons in the ground state, and $V_\text{ext}$ is the external potential; the fourth term in the right-hand-side (RHS) of Eq.~(\ref{eq:gnlGPE}) represents the higher-order correction to the local density approximation and will be discussed later together with the meaning of nondimensional constants $\chi$ and $\gamma$. The third, integral term in the RHS describes \textit{nonlocal} interatomic interactions determined by the two-body, isotropic, long-range potential $V(r)$, where $r=\vert\xx-\xx'\vert$.

Perturbing Eq. \eqref{eq:gnlGPE} around equilibrium leads to the generalized Bogoliubov dispersion relation
\begin{equation}
    \omega(\kk) = c k \sqrt{\frac{\xi^2 k^2}{2} + \frac{\hat{V}(\kk) + \chi (\gamma+1)}{1+\chi (\gamma+1)}}\,,
    \label{eq:dispersion_relation}
\end{equation}
where $\kk$ is the wave vector of the perturbation, and
\begin{equation}
c = \sqrt{\frac{g n_0f_{\rm NL}}{m_4} }\,, \,\, \xi = \frac{\hbar}{\sqrt{2m_4gn_0 f_{\rm NL}}}\,, \,\, f_{\rm NL} = 1 +\chi(\gamma+1)
\label{eq:cxiV}
\end{equation}
are the speed of sound and the healing length, respectively. The Fourier transform of the interaction potential $\hat{V}(\kk) =  \int e^{i\kk \cdot \xx} V(\xx)\,d\xx$ is normalized such that $\hat{V}(k=0) = 1$. Note that with this convention $\omega(k)\approx c k$ for $k\to 0$.

Assuming the locality of interactions between bosons, so that the potential $V$ becomes the $\delta$ function, $V(\vert\xx-\xx'\vert)=\delta(\xx-\xx')$, and neglecting the higher-order correction ($\chi=0$), Eq.~(\ref{eq:gnlGPE}) reduces to the classical, (\textit{``standard''}) Gross-Pitaevskii equation (GPE) which provides an accurate quantitative description of dilute quantum gases but, in general, cannot serve as a quantitative model of dense \4He. Note, however, that the classical GPE gives a good qualitative (and in some cases even quantitative) description of the vortex nucleation by moving objects in liquid helium at temperatures close to absolute zero, see, e.g., Refs. \cite{Huepe97,Rica2001,Winiecki99,Winiecki2000,Pham2005,Stagg16,Stagg17,Frisch24}. Of particular relevance to our work is the numerical study by Villois \etal \cite{VilloisSalman} of the electron bubbles' mobility limited by vortex nucleation. However, the standard GPE is not applicable to the analysis of roton emission simply because the excitation spectrum for this equation is strictly monotonic and therefore cannot reproduce the Landau spectrum of excitations in \4He.

An idea of using the long-range potential $V(r)$ for a phenomenological description of superfluid helium was discussed by Gross \cite{Gross63} and later by Dupont-Roc \etal \cite{Dupont-Roc90}, Pomeau and Rica \cite{Pomeau93}, and Dalfovo \etal \cite{Dalfovo95}. It is now well understood \cite{Berloff1999,BerloffRobertsVI-1999,Berloff2014,Reneuve2018,Muller20,Muller22} that with an appropriate choice of parameters modeling the interaction potential $V(r)$ the excitation spectrum for the gnlGPE (\ref{eq:gnlGPE}) can approximate fairly accurately the phonon, maxon, and roton parts of the Landau excitation spectrum of \4He.

Based on the density functional theory \cite{Dupont-Roc90,Dalfovo95} for \4He, the higher-order correction term, $g\chi n_0^{-\gamma}\vert\psi\vert^{2(1+\gamma)}$ was added by Berloff and Roberts \cite{BerloffRobertsVI-1999} when they realized \cite{Berloff1999} (see also Ref. \cite{Reneuve2018}) that the replacement alone of the local interatomic interaction potential $V(\vert\xx-\xx'\vert)=\delta(\xx-\xx')$ in the classical GPE with the nonlocal one leads, in the numerical analysis, to the emergence of nonphysical mass concentrations. We note here that, together with a choice of potential $V(r)$, a choice of the constants $\chi$ and $\gamma$ makes possible to reproduce the equation of state for liquid helium. In particular, the value $\gamma=2.8$ is required to reproduce \cite{BerloffRobertsVI-1999,Muller20} the long-wave speed of sound which is proportional to $\rho^{2.8}$. 

Nonlocal Gross-Pitaevskii equations, with a spectrum of excitation possessing a roton-like minimum, also emerge in the theory of dipolar quantum gases (see, e.g., Refs. \cite{Lahaye2009,Parker2008}). Among very recent works addressing the issue of roton's nature, it is worth mentioning an interesting theoretical study~\cite{Villois2025} of elementary excitations in two-dimensional dipolar Bose-Einstein condensates. In full agreement with Feynman's hypothesis, the authors of Ref.~\cite{Villois2025} found that quantized vortex dipoles which, in 2D, model three-dimensional quantized vortex rings, transit into roton excitations rather than into phonons.

Based on the gnlGPE (\ref{eq:gnlGPE}), some of us investigated theoretically and numerically \cite{Muller22}, in the zero-temperature limit, the mechanisms of roton emission and of nucleation of quantized vortices by objects moving in superfluid \4He and calculated the associated critical velocities (see also a recent review \cite{SkrbekSergeevCritVel} of this and related works). In the cited work \cite{Muller22}, for the interatomic interaction potential $V(r)$ we chose an analytical form such that the dispersion relation derived from Eq.~(\ref{eq:gnlGPE}) approximates, fairly accurately, the Landau excitation spectrum \cite{GodfrinDispersion} at zero pressure (see Fig.~\ref{fig:Godfrin}). 

In what follows, the mechanisms of superfluidity breakdown will be studied numerically, and the associated critical velocities calculated, in the two-dimensional setting, for interaction potentials $V(r)$ which correspond to the dispersion curves that approximate the spectra of excitations in \4He at elevated pressures in the range from 0 to 24~bar shown in Fig.~\ref{fig:RotonMin}. Instead of using an analytical expression for $V(r)$, the potentials are obtained by inverting Eq.~\eqref{eq:dispersion_relation} using explicitly the experimental data of the dispersion relation reported in \citep{GodfrinDispersion} to obtain directly $\hat{V}(k)$ (see Appendix~\ref{App:nonlocalpotential} for more details). In this way, our model satisfies by construction the observed helium dispersion relation at different pressures.

\section{Results}

\label{sec:results}

To obtain the critical velocity for the breakdown of superfluidity, we analyze the stationary solutions for the superfluid in the frame of reference comoving with the obstacle. To do so, we perform numerical simulations of the gnlGPE \eqref{eq:gnlGPE} in two dimensions using the pseudo-spectral code FROST \cite{Krstulovic2020}, in a square of size $L=2\pi$ with periodic boundary conditions. We study resolutions that vary from $N^2 = 128^2$ to $N^2 = 6400^2$ linear collocation points, fixing the healing length $\xi = 1.5 \Delta x$ with $\Delta x = L/N$ the resolution, $c=1$ and $n_0=1$ (except for $P=0$ where we fix $\xi=3 \Delta x$ to solve the Pitaevskii plateau). We vary the size of the moving obstacle from $D=10 \xi$ to $D=1000 \xi$, keeping always $L>5 D$ to avoid spurious effects arising from the boundary conditions. The obstacle in two dimensions is imposed using an external potential with a Gaussian profile
\begin{equation}
 \Vext(\xx - \Uvec t) = V_0 \exp\left(-\frac{|\xx - \Uvec t|^2}{2\Delta^2}\right)\,.
 \label{eq:VGauss}
\end{equation}
In the Thomas-Fermi approximation, the size of the obstacle is determined by the relation $\Delta = D/[2\sqrt{2\log(V_0)}$], with $D$ being the diameter of the disk, and the amplitude of the potential $V_0 = 10 \gg 1$ to completely deplete the superfluid. 

\begin{figure*}%[h]
\centerline{\includegraphics[width=.9\linewidth]{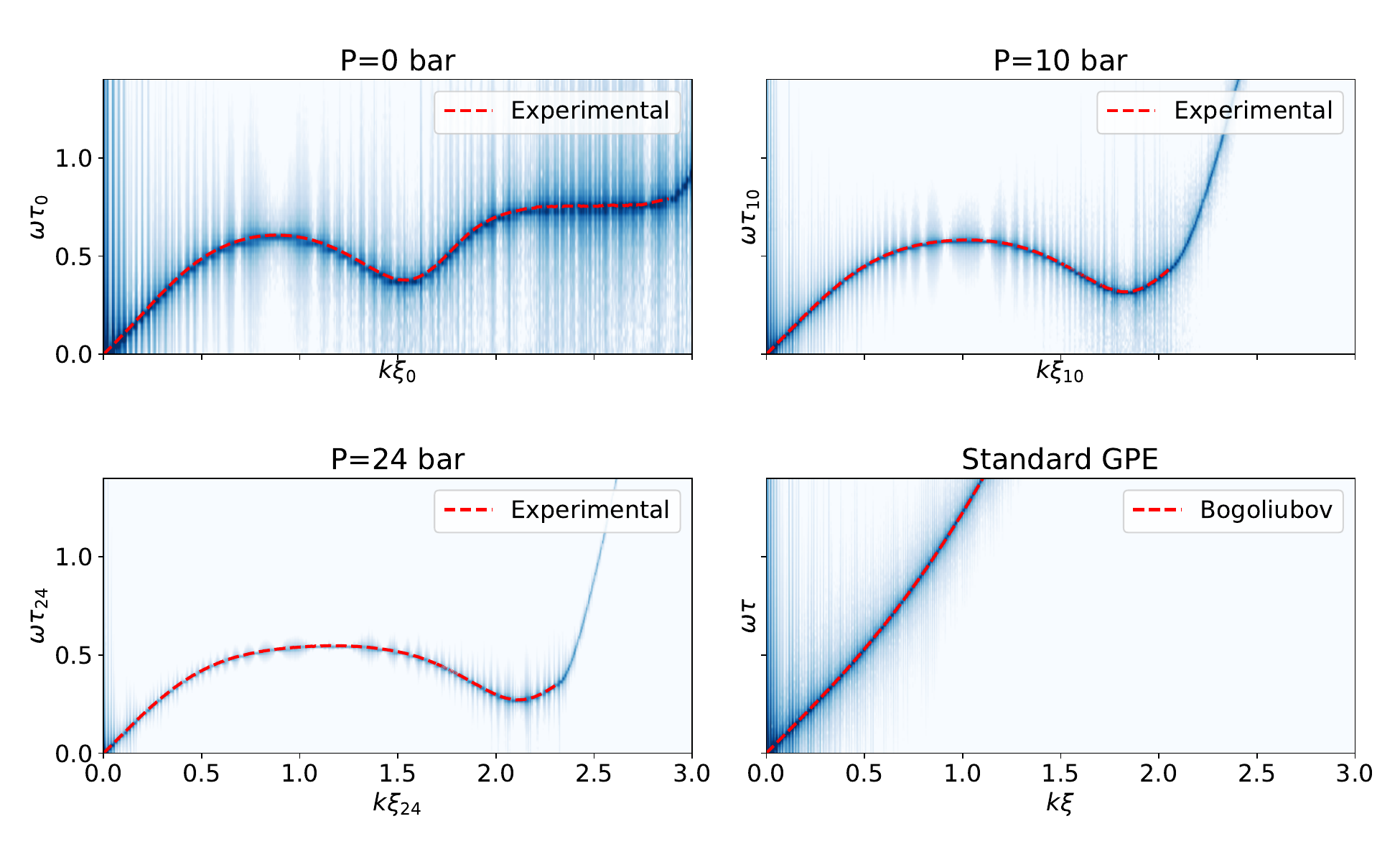}}
\medskip
\caption{Dispersion relation of the gnlGPE for interaction potentials for superfluid $^4$He at different pressures. For comparison, we also show the Bogoliubov dispersion relation for the standard GPE.
}
\label{fig:dispersion}
\end{figure*}

Assuming the steady wave solutions in the form $\psi(\xx - \Uvec t)$, we then look for maxima or minima of the free energy using a Newton-Raphson method, coupled with an iterative bi-conjugate gradient stabilized method to improve convergence (see \cite{Muller22,Tuckerman2004} for details on the methods). Varying the Mach number, defined as $M=U/c$ (here $U=\vert\Uvec\vert$), and the diameter $D$ of the obstacle, we characterize the bifurcation diagram for the energy of stable and unstable solutions. Stable solutions typically correspond to the moving obstacle, while unstable solutions may have pair of quantum vortices that move along with the obstacle \cite{Huepe97,Pham2005,Muller22}. The maximum Mach number obtained via this analysis determines the critical velocity for superfluidity, defined as $M^\text{crit} = U^\text{crit}/c$. Velocities larger than this one do not correspond anymore to stationary solutions, and elementary excitations start to be generated in the wake of the obstacle.

\begin{table}[h]
\centering
\begin{tabular}{c|c|c|c|c|c}
    $P$ [bar] & $c$ [m/s] & $\xi$ [\angstrom] & $\krot$ [\angstrom$^{-1}$] & $\vLc$ [m/s] & $\MLc$ \\
    \hline
    0   & $238.3$ & 0.8  & 1.918 &  58.76 & 0.247 \\ 
    2   & $253.9$ & 0.83  & 1.935 &  57.37 & 0.229 \\ 
    5   & $274.0$ & 0.85  & 1.957 &  55.49 & 0.202 \\ 
    10  & $302.3$ & 0.92 & 1.988 &  52.69 & 0.174 \\ 
    24  & $361.9$ & 1.03 & 2.048 &  46.45 & 0.128
\end{tabular}
\caption{Dependence of the speed of sound $c$ and the healing length $\xi$ on the pressure $P$. Values of $\xi$ are extracted from \citet{Donnelly1991}, while the rest are extracted from \citet{GodfrinDispersion}. 
\label{tab:helium}}
\end{table}

We impose an interaction potential between bosons that describes the experimental dispersion relation observed in He II according to Ref.~\cite{GodfrinDispersion} for pressures $P=0$, $2$, $5$, $10$, and $24$~bar. Note that not only the dispersion relation changes with pressure in He II, but also the speed of sound $c$ and the healing length $\xi$. Their values are shown in Table~\ref{tab:helium} \cite{GodfrinDispersion,Donnelly1991}. {For pressures $P=0$, $10$, and $24$~bar the dispersion relation $|\hat{\psi}(k,\omega)|^2$ is illustrated in Fig.~\ref{fig:dispersion} and compared with that for the standard GPE. For wave vectors larger than the experimental accessible values, we adjust the interaction potential to make the dispersion curve go smoothly towards the $k^2$-`free-particle' dispersion relation of the Gross-Pitaevskii model.
The Pitaevskii plateau was only measured experimentally for $P=0$ bar, and to solve it properly we increase the resolution of the simulations only for this case. For each pressure, the wave vector is normalized by the pressure dependent healing length $\xi$, and the frequency is normalized by the vortex time scale $\tau_P = \xi/ c$, where $c$ is the pressure dependent speed of sound; $\xi=\xi(P)$ and $c=c(P)$ are recorded in Table~\ref{tab:helium}.

The main results of this article are summarized in Fig.~\ref{fig:critical_mach}.
\begin{figure}
\includegraphics[width=.49\linewidth]{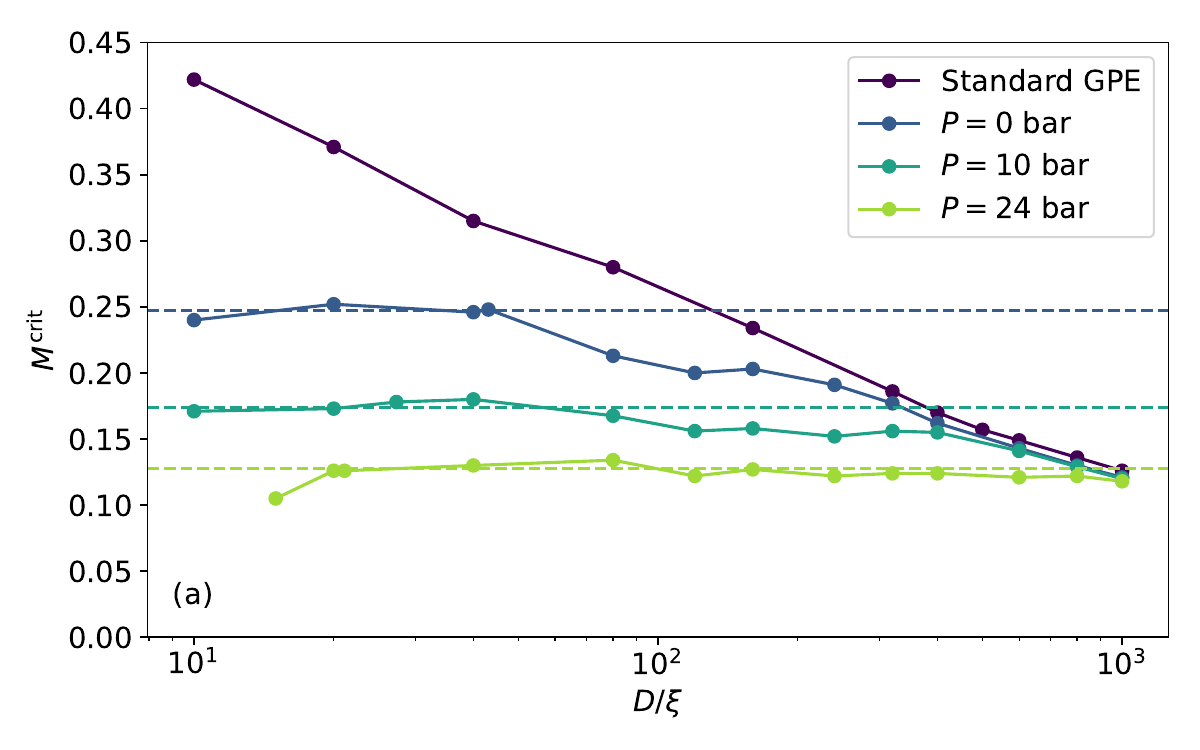}
\includegraphics[width=.49\linewidth]{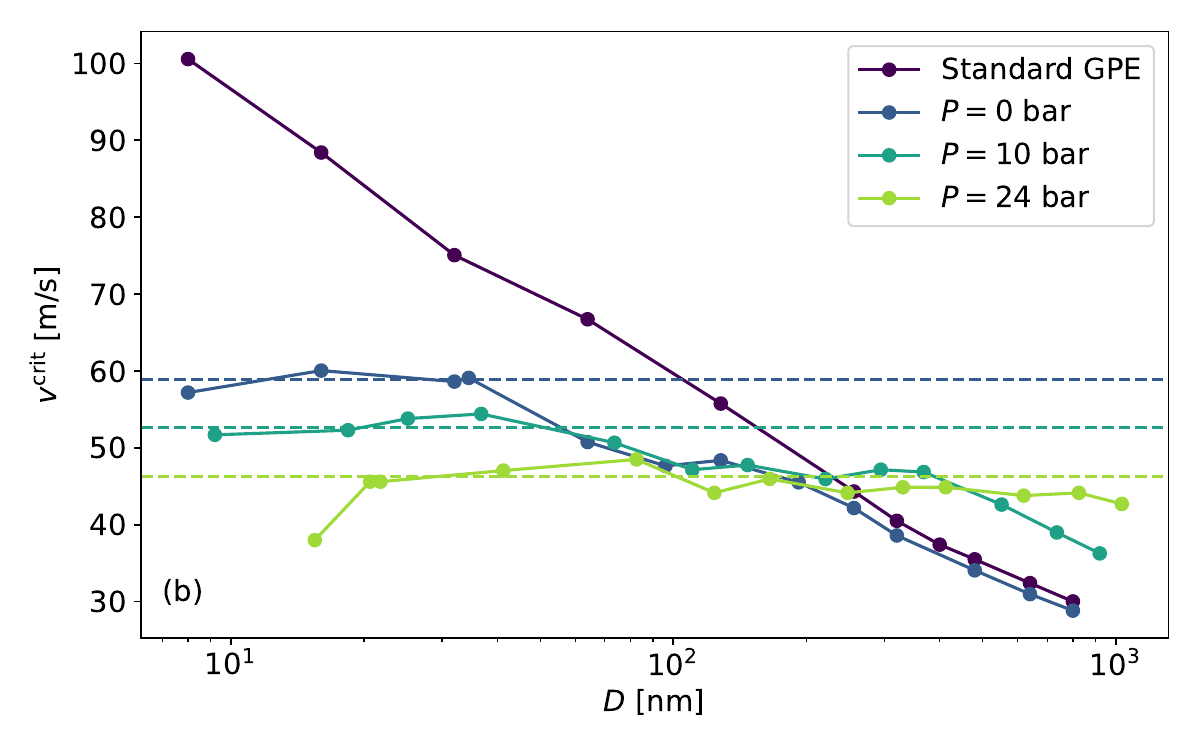}
\caption{(a) Critical Mach number $\Mc = \vc / c$ as a function of the obstacle's diameter $D$ normalized by the healing length $\xi$ at different pressures. Horizontal dashed lines indicate Landau's critical Mach number $\MLc$. Panel (b) shows the same curves in physical units.
}
\label{fig:critical_mach}
\end{figure} 
Landau's criterion for superfluidity Eq.\eqref{eq:LandauCriterium} states that the critical velocity of the system is given by $\vLc = \min_k \omega(k)/k=\Delta_\text{R}/\hbar k_0$  (see Table~\ref{tab:helium}). This critical velocity decreases as the pressure increases, and when normalized with the speed of sound (that also increases with pressure), the Mach number decreases even faster. For instance, at $P=0$ Landau's Mach number is $\MLc = 0.247$, while at $P=24$ bar $\MLc = 0.128$, showing a difference factor close to $2$. For obstacles of small sizes, the critical Mach for superfluidity is determined by this value. The other critical Mach number, $\Mvc=\vvc/c$, where $\vvc$ is the critical velocity of nucleation of quantized vortices, is obtained from the solution of the standard GPE. This critical Mach number decreases with the obstacle's size. For all pressures, as the size of the obstacle increases, the critical Landau's Mach numbers also start to decrease, and eventually they all seem to collapse on the single curve determined by the Mach number of vortex nucleation, $\Mvc$. For larger pressures, the critical Mach number for rotons extends to larger obstacles, and the gap between the roton and vortex Mach numbers becomes larger. This result suggests that it might be simpler to detect rotons at larger pressures. Figure \ref{fig:critical_mach}b shows the same curves using physical units. For particles of small sizes, $D < 100$ nm, there is a gap between the critical velocity for rotons and the one of vortices (obtained using the standard GPE) that increases with pressure.

In He II, the dynamics of small particles is studied through negative ions, accelerated by an external electric field to a certain velocity. Typical values of the size of these bubbles for different pressures are shown in Table~\ref{tab:ions}. 
\begin{table}[h]
\centering
\begin{tabular}{c|c|c|c|c|c}
    $P$ [bar] & $0$ & $2$ & $5$ & $10$ & $24$ \\
    \hline
    $\xi$ [\angstrom]   & $0.8$ & $0.83$  & $0.85$ &  $0.92$ & $1.03$ \\ 
    $D$ [\angstrom]  & $34.5$ & $30.6$ & $27.8$ &  $25.2$ & $21.7$ \\ 
    $D/\xi$  & $43.1$ & $36.9$ & $32.7$ &  $27.3$ & $21.1$ \\
    $c$ [m/s] & $238.3$ & $253.9$ & $274.0$ & $302.3$ & $361.9$
\end{tabular}
\caption{Dependence of the healing length $\xi$, the speed of sound $c$ and negative ion diameter $D$ on the pressure $P$. Values of $\xi$ are extracted from \citet{Donnelly1991}, values of $c$ are extracted from \citet{GodfrinDispersion}, and $D$ are extracted from \cite{Ostermeier}. \label{tab:ions}}
\end{table}
From now on, the sizes of the obstacles we use in the simulations at different pressures correspond to the ion sizes. Figure \ref{fig:ion_critical_velocities}
\begin{figure}
\centerline{\includegraphics[width=.99\linewidth]{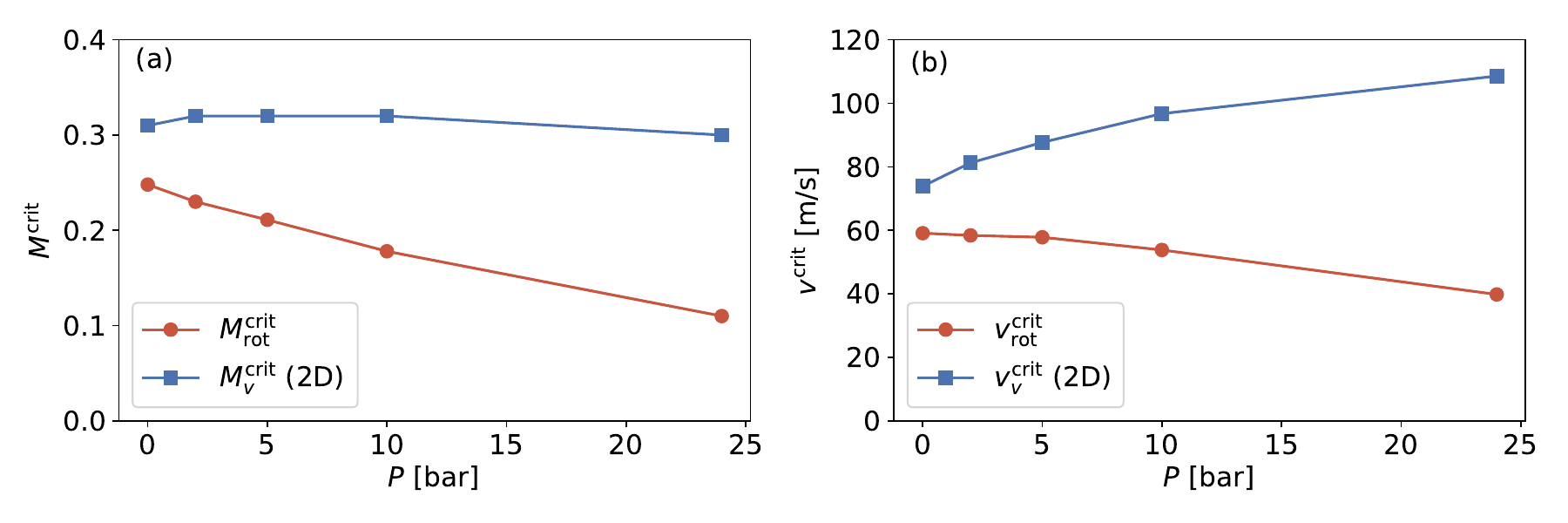}}
\caption{Critical (a) Mach numbers and (b) velocities for roton emission ($\Mrc$ and $\vrc$, respectively) and vortex nucleation ($\Mvc$ and $\vvc$) for negative ions at different pressures. The sizes of the ions are reported in Table~\ref{tab:ions}.
}
\label{fig:ion_critical_velocities}
\end{figure}
shows the critical Mach numbers and velocities of these ions at different pressures. The roton critical velocity is obtained using the same minimization method as the one used to obtain Fig.~\ref{fig:critical_mach}, while the vortex nucleation critical velocity is obtained using a dynamic approach, in which the particle is accelerated until it reaches a certain target velocity; we let it evolve for a time around $10 \xi/c$ and look if any vortex was nucleated. We do this using steps of $\Delta M=0.01$. 
Note that with this method, vortices nucleate only when rotons have already been emitted. Supplemental simulations which include dissipation confirm that rotons do not significantly affect the critical velocity for vortex nucleation. This supports our treatment of roton emission and vortex nucleation as independent processes under prescribed velocity conditions.
Figure \ref{fig:ion_critical_velocities}b shows that the critical velocity for emission of rotons decreases with pressure, while the one for vortex nucleation increases, trend that qualitatively agrees with experimental observations, e.g. \cite{Ellis85,McClintockBowleyProgressLT95}. As the gap $\dvc = \vvc - \vrc$ increases, there is a larger region in which rotons could be observed in the absence of quantum vortices.
To confirm this picture, Fig.~\ref{fig:visualization}
\begin{figure*}
\centerline{\includegraphics[width=.99\linewidth]{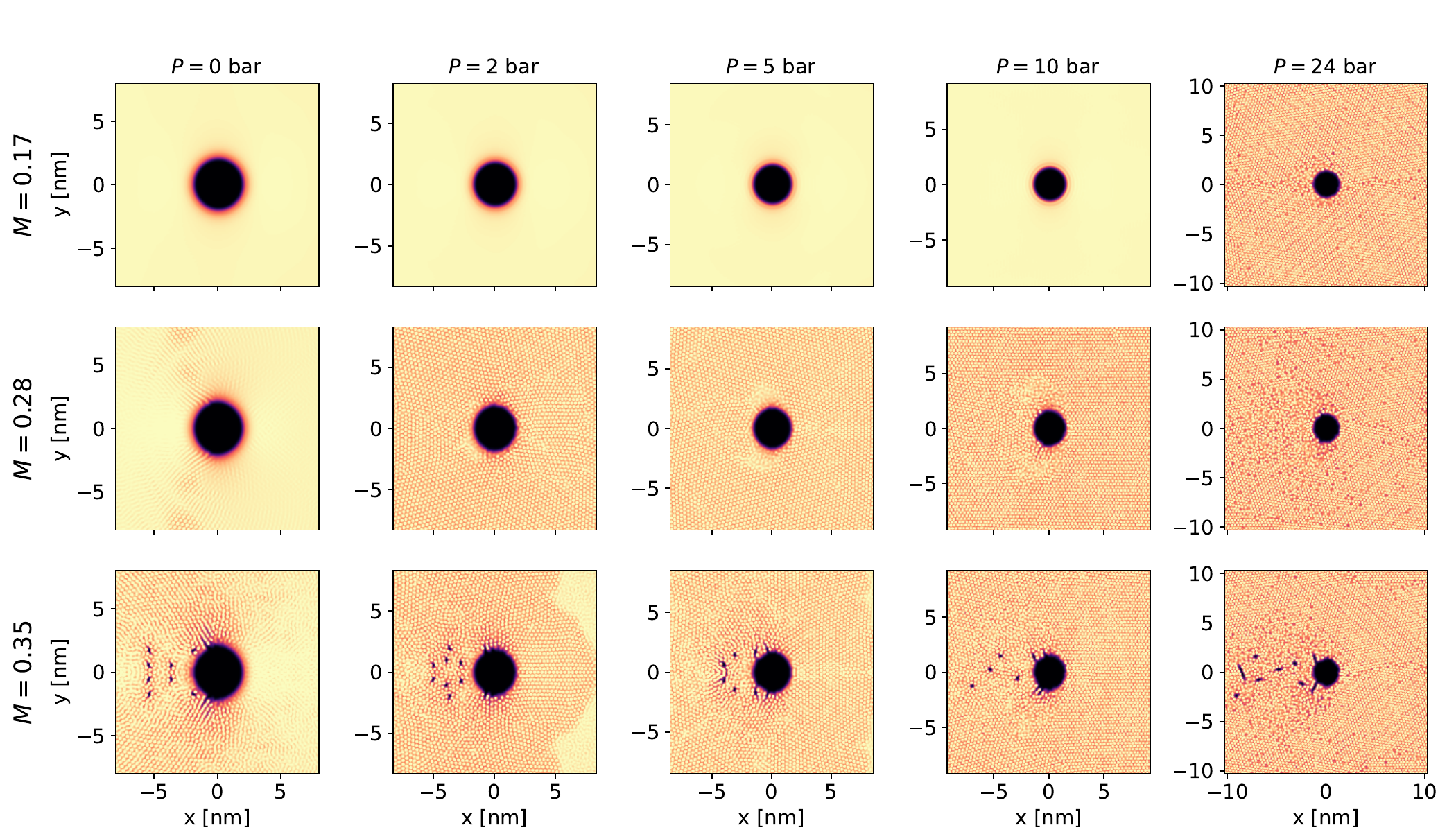}}
\caption{Mass density of the superfluid at different pressures for negative ions moving at different Mach numbers. Values of $M$ are chosen to illustrate the excitation of rotons and quantum vortices.}
\label{fig:visualization}
\end{figure*}
shows a visualization of the mass density of the superfluid in the presence of ions moving at different Mach numbers. For $M=0.17$, only the moving ion at $P=24$ bar excites rotons in the system. For $M=0.28$, rotons are emitted for all pressures, and we observe no nucleation of quantum vortices, as its critical Mach number is around $\Mvc \approx 0.3$. For the largest Mach number shown here, $M=0.35$, we see that vortices are nucleated in all cases.
}

To confirm that the honeycomb-like structure observed in Fig.~\ref{fig:visualization} correspond to rotons, we analyze the mass density spectrum of the superfluid, $|\hat{\psi}(k_x,k_y)|^2$ (here $\hat{\psi}$ is the two-dimensional Fourier transform of the macroscopic wave function). Figure \ref{fig:visualization_rotons} shows that for all velocities, low wave numbers are excited due to the density depletion introduced by the moving obstacle. Below Landau's critical velocity, these are the only excited wave numbers in the system. As the velocity exceeds Landau's critical one, wave numbers close to $\krot$ become excited, confirming the relation between the honeycomb-like structures and rotons. Finally, for higher velocities, more wave numbers get excited due to non-linear interactions and the presence of vortices.

\begin{figure*}
\centerline{\includegraphics[width=.99\linewidth]{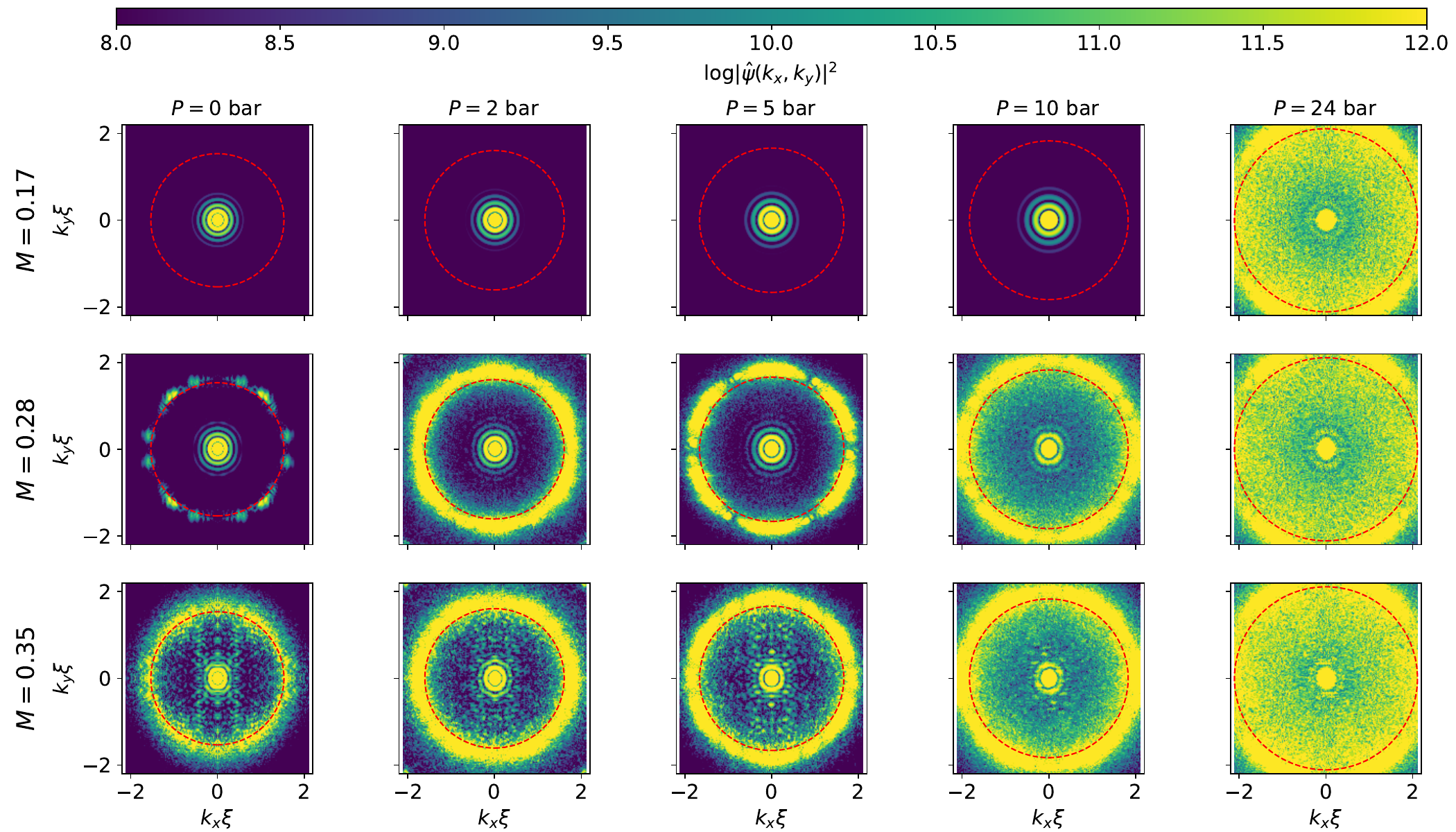}}
\caption{Two-dimensional Fourier transform of the mass density $|\hat{\psi}(k_x,k_y)|^2$ for the same times as in Fig.~\ref{fig:visualization}. Red dotted lines indicate the wave number $\krot$ associated with rotons for each pressure.} 
\label{fig:visualization_rotons}
\end{figure*}

\section{Conclusions}

We studied the critical velocities for vortex nucleation and roton emission in superfluid $^4$He at pressures ranging from $0$ to $24$ bar. We performed numerical simulations of the generalized, nonlocal Gross-Pitaevskii equation, imposing a nonlocal interaction potential that reproduces the excitation spectrum observed experimentally \cite{GodfrinDispersion} in superfluid $^4$He across this pressure range. By introducing a moving object of varying sizes and velocities, we identified the critical velocity for roton emission through the characterization of the bifurcation diagram via a minimization method, and a second critical velocity for vortex nucleation through a dynamic approach. 

As pressure increases, the critical Mach number for rotons decreases from $\Mrc \approx 0.247$ at $0$ bar to $\Mrc \approx 0.128$ at $24$ bar, values that are close to Landau critical velocity. This behavior is observed for small particles, where all curves display different critical velocities. As the size of the particle reaches values $D \geq 1000 \xi$, the curves for all pressures collapse following the same behavior as for the standard GPE, associated with the critical velocity for vortex nucleation. 

Focusing on the dynamics of negative ions, we found that the critical velocity for vortex nucleation increases with pressure, while the critical roton velocity decreases. These results are qualitatively in agreement with experimental observations in superfluid $^4$He \cite{Ellis85,McClintockBowleyProgressLT95}.
The gap between the critical velocities for roton emission and vortex nucleation increases with pressure, suggesting that detection of rotons, in the absence of vortices, could be achieved at high pressure. 
The critical velocity for vortex nucleation is not significantly affected by the presence of rotons. In our approach, a fixed velocity of the moving obstacle is prescribed; this justifies an independent treatment of the two critical velocities. However, such a treatment would no longer hold if the drag force on the obstacle were computed. Such a scenario lies beyond the scope of this work and is left for future investigation.

Because of the numerical complexity of our study, we have restricted ourselves to two-dimensional (2D) simulations. Remarkably, the numerical values obtained from our simulations are comparable to those reported in experiments. However, a detailed study of the critical velocity in three dimensions (3D) remains computationally prohibitive, particularly for determining the vortex configuration at the onset of nucleation.
Analytical calculations by Muirhead, Vinen, and Donnelly \cite{Muirhead}, using a pure hydrodynamical approach, predict that vortex loops—vortices with both ends attached to the ion—are more likely to nucleate than a vortex ring located at the ion’s equator. In contrast, numerical works using the standard Gross–Pitaevskii equation (GPE) in 2D \cite{Huepe97} and 3D \cite{Winiecki99} show that, while an ion–vortex loop is more energetically favorable than an ion–vortex ring configuration, the energy gap between the two reduces as the ion’s speed increases, eventually vanishing at the critical velocity. Thus, in practice, both configurations may coexist once vortices begin to nucleate. What constitutes the \emph{minimal seed} of vortex nucleation in 3D remains an interesting and open question, particularly in the context of the generalized Gross-Pitaevskii equation, and should be an object of future investigation. 
Finally, it is important to note that, in our simulations, the ion is modeled as a perfectly spherical hard sphere. Varying the ion modeling—for example, by considering a bubble-like object \cite{VilloisSalman} or non-spherical ions—could modify the measured values, though trends with pressure may remain unaffected.

To conclude, we demonstrated that the pressure-dependent physical mechanisms and associated critical velocities of roton emission and vortex nucleation by the moving objects in zero-temperature superfluid \4He can be analyzed within a single theoretical framework based on the generalized, nonlocal Gross-Pitaevskii model.

\acknowledgements

G.K. acknowledges support
of Agence Nationale de la Recherche through the project
QuantumVIW ANR-23-CE30-0024-02. LS acknowledges support of Czech Science Foundation (GACR) under \#25-16588S. This project was provided with computer and storage resources by GENCI at CINES thanks to the grant A0200517369 on the supercomputer Adastra GENOA.

\appendix

\section{Non-local potential\label{App:nonlocalpotential}}
The choice of the non-local potential in the gnlGPE \ref{eq:gnlGPE}, and more particularly its Fourier transform $\hat{V}({\bf k})$, directly determines the dispersion relation through Eq.~\eqref{eq:gnlGPE}. In previous approaches, the shape of the non-local potential was chosen by adjusting parameters of a fit such that the gnlGPE is close to reported experimental values for superfluid helium \cite{Reneuve2018,Muller20,Muller22}. In this work, we directly use recent experimental data furnished in reference \cite{GodfrinDispersion}. We proceed as follows.

We first note that by inverting Eq.~\eqref{eq:gnlGPE}, the non-local potential reads
\begin{equation}
    \hat{V}(k)=\left(\frac{\omega(k)^2}{c^2k^2}- \frac{\xi^2k^2}{2}\right)[1+\chi(1+\gamma)] - (1 +\gamma)\chi. \label{eq:Vkdata}
\end{equation}
In order to use the experimental data, we express the experimental values of the dispersion relation $\omega^{\rm exp}(k)$ in dimensionless units using the speed of sound and the vortex core size at each pressure. Then, $\omega^{\rm exp}(k)$ is  interpolated using a cubic BSpline in a much finer grid of $5000$ points between $k=0$ and $k_{\rm max}^{\rm exp}$, the largest available experimental wave vector. With this procedure, we could directly input the non-local potential in our gnlGPE code and reproduce, by construction, the experimental dispersion relations displayed in Fig.~\ref{fig:dispersion}.
However, in order to correctly resolve vortices, we need values of the non-local potential for very large wavevectors ($k\xi\gg1$ ), which are not available experimentally. In order to properly allow for a $k^2$-\emph{free-particle} smooth dispersion relation at high $k$, we extend $\hat{V}^{\rm exp}(k)$ from $\xi k_{\rm max}^{\rm exp}$ up to $k_{\max} \xi = 7$ for $P=0$ bar and $\xi k_{\max}=4$ for $P>0$. We impose a smooth quadratic prolongation of $\hat{V}(k)$ from $k_{\rm max}^{\rm exp}$ to $k_{\max}$, where it vanishes. We then use the generated $\hat{V}^{\rm exp}(k)$ in the whole range of our simulations leading to the data displayed in Fig.~\ref{fig:dispersion}.


\begin{thebibliography}{99}

\bibitem{Kapitza}
P. Kapitza, Viscosity of Liquid Helium below the $\lambda$-Point, Nature \textbf{141}, 74 (1938). 

\bibitem{Allen}
J. F. Allen and A. D. Misener, Flow of Liquid Helium II, Nature \textbf{141}, 75 (1938).

\bibitem{London}
F. London, The $\lambda$-Phenomenon of Liquid Helium and the Bose-Einstein Degeneracy, Nature \textbf{141}, 643 (1938).

\bibitem{Andronikashvili}
E. Andronikashvilli, A direct observation of two kinds of motion in helium II, J. Phys. (USSR) {\bf 10}, 201 (1946).

\bibitem{Tisza}
L. Tisza, Transport Phenomena in Helium II, Nature \textbf{141}, 913 (1938); Sur la superconductibilit\'e thermique de l’h\'elium~II liquide et la statistique de Bose-Einstein, C. R. Acad. Sci. \textbf{207}, 1035 (1938); La viscosit\'e de l’h\'elium II liquide et la statistique de Bose–Einstein, C. R. Acad. Sci. \textbf{207}, 1186 (1938).

\bibitem[Landau(1941)]{Landau41}
L. D. Landau, The theory of superfluidity of Helium II,  J. Phys. (USSR) {\bf 5}, 71 (1941). 

\bibitem[Landau(1947)]{Landau47}
L. D. Landau, On the theory of superfluidity of Helium II,  J. Phys. (USSR) {\bf 11}, 91 (1947).

\bibitem{Peshkov1}
V. Peshkov, Second Sound in Helium II, J. Phys. (USSR) {\bf 8}, 381  (1944). 

\bibitem{Peshkov2}
V. P. Peshkov, Determination of the Velocity of Propagation of the Second Sound in Helium II, J. Phys. (USSR) \textbf{10}, 389 (1946). 

\bibitem{Onsager}
L. Onsager, In discussion on paper by C. J. Gorter, Nuovo Cimento \textbf{6}, suppl. 2, 249 (1949). 

\bibitem{VinenSingle}
W. F. Vinen, Detection of single quanta of circulation in liquid helium II, Proc. R. Soc. London A \textbf{260}, 218 (1961).

\bibitem{Feynman}
R. P. Feynman, Application of quantum mechanics to liquid helium, in \textit{Progress in Low-Temperature Physics}, edited by C.~J.~Gorter, Vol.~1 (North Holland, 1955), pp.~17–53.

\bibitem{VinenOld}
W. F. Vinen, Mutual friction in a heat current in liquid helium II, Proc. R. Soc. A \textbf{240},
114 (1957); \textbf{240}, 128 (1957); \textbf{242}, 493 (1957); \textbf{243}, 400 (1958).

\bibitem{FeynmanCohen56}
R. P. Feynman and M. Cohen, Theory of Inelastic Scattering of Cold Neutrons from Liquid Helium, Phys. Rev. \textbf{102}, 1189
(1956).

\bibitem[Godfrin \textit{et al.}(2021)]{GodfrinDispersion}
H. Godfrin, K. Beauvois, A. Sultan, E. Krotscheck, J. Dawidowski, B. F{\aa}k, and J. Ollivier, 
Dispersion relation of Landau elementary excitations and thermodynamic properties
of superfluid $^4$He,  Phys. Rev. B {\bf 103}, 104516 (2021).

\bibitem{Glyde2018}
H. R. Glyde, Excitations in the quantum liquid $^4$He: A review, Rep. Prog. Phys. \textbf{81}, 014501 (2018).

\bibitem{Meyer61}
L. Meyer and F. Reif, Ion Motion in Superfluid Liquid Helium under Pressure, Phys. Rev. \textbf{123}, 727–731 (1961).

\bibitem{RR64}
G. W. Rayfield and F. Reif, Quantized Vortex Rings in Superfluid Helium,  Phys. Rev. \textbf{136}, A1194 (1964).

\bibitem{Rayfield66}
G. W. Rayfield, Roton emission from negative ions in helium~II, Phys. Rev. Lett. \textbf{16}, 934 (1966).

\bibitem{SkrbekSergeevCritVel}
L. Skrbek and Y. A. Sergeev, Critical velocities in flows of superfluid $^4$He, Phys. Fluids \textbf{37}, 031305 (2025).

\bibitem{Phillips73}
A. Phillips and P. V. E. McClintock, Evidence for roton creation in a superfluid field emission diode near 0.5~K, Phys. Lett. A \textbf{46}, 109 (1973).

\bibitem{Allum77}
D. R. Allum, P. V. E. McClintock, A. Phillips and R. M. Bowley, The breakdown of superfluidity in liquid $^4$He: an experimental
test of Landau’s theory, Philos. Trans. R. Soc. London A {\bf 284}, 179 (1977).

\bibitem[Stamp(1979)]{Stamp79}
 P. C. E. Stamp, P. V. E. McClintock, and W. M. Fairbairn, Possible influence of thermal rotons on vortex nucleation by negative ions in pressurised He~II below 1~K,  J. Phys. C: Solid State Phys. {\bf 12}, L589 (1979).
 
\bibitem{Nancolas82}
G. G. Nancolas and P. V. E. McClintock, Quenching of the lon/Vortex-Ring Transition in He~Il by Intense Electric Fields, Phys. Rev. Lett. \textbf{48}, 1190 (1982).

\bibitem{Ellis82}
T. Ellis and P. V. E. McClintock, Effective Mass of the Normal Negative-Charge Carrier in Bulk He~II, Phys. Rev. Lett. \textbf{48}, 1834 (1982).

\bibitem{Bowley82}
R. M. Bowley, P. V. E. McClintock, F. E. Moss, G. G. Nancolas, and P. C. E. Stamp, The breakdown of superfluidity in liquid $^4$He: III. Nucleation of quantized vortex rings,
Philos. Trans. R. Soc. London A {\bf 307}, 201 (1982).

\bibitem[Ellis and McClintock (1985)]{Ellis85}
T. Ellis and P. V. E. McClintock, The breakdown of superfluidity in liquid 4He. V. Measurement of the Landau critical velocity for roton creation, Philos. Trans. R. Soc. London A {\bf 315}, 259 (1985).

\bibitem{Nancolas85-PhilTrans}
G. G. Nancolas, R. M. Bowley, and P. V. E. McClintock, The breakdown of superfluidity in liquid $^4$He. IV. Influence of $^3$He isotopic impurities on the nucleation of quantized vortex rings, Philos. Trans. R. Soc. London A {\bf 313}, 537 (1985).

\bibitem{Nancolas85}
G. G. Nancolas, T. Ellis, P. V. E. McClintock, and R. M. Bowley, A new form of energy dissipation by a moving object in He~II, Nature \textbf{316}, 797 (1985).

\bibitem{Hendry88}
P. C. Hendry, N. S. Lawson, P. V. E. McClintock, C. D. H. Williams, and R. M. Bowley, Macroscopic Quantum Tunneling of Vortices in He~II, Phys. Rev. Lett. \textbf{60}, 604 (1988).

\bibitem{Hendry90}
P. C. Hendry, N. S. Lawson, P. V. E. McClintock, C. D. H. Williams, and R. M. Bowley, The breakdown of superfluidity in liquid $^4$He. VI. Macroscopic quantum tunnelling by vortices in isotopically pure He~II, Philos. Trans. R. Soc. London A {\bf 332}, 387 (1990).

\bibitem[McClintock and Bowley(1995)]{McClintockBowleyProgressLT95}
P. V. E. McClintock and R. M. Bowley, The Landau critical velocity,  in \textit{Progress in Low Temperature Physics}, edited by W. P. Halperin (Elseivier, 1995), Vol. XIV,  pp.~1–68.

\bibitem{Bowley77}
R. M. Bowley and F. W. Sheard, Motion of negative ions at supercritical drift velocities in liquid $^4$He at low temperatures, Phys. Rev. B \textbf{16}, 244 (1977).

\bibitem{Bowley84}
R. M. Bowley, Nucleation of vortex rings by negative ions in liquid
helium at low temperatures, J. Phys. C: Solid State Phys. \textbf{17}, 595 (1984).

\bibitem{Muirhead}
C. M. Muirhead, W. F. Vinen, and R. J. Donnelly, The nucleation of vorticity in superfluid $^4$He. I.
Basic theory, Philos. Trans. R. Soc. Lond. A \textbf{311}, 433 (1984).

\bibitem{BarenghiMcClintockMuirhead}
C. F. Barenghi, P. V. E. McClintock, and C. M. Muirhead, Vinen’s Energy Barrier, J. Low Temp. Phys.
\textbf{212}, 185 (2023).

\bibitem{Iordanskii68}
S. V. Iordanski\u{\i}, Calculation of the critical current of extraneous particles in a Bose system, Soviet Phys. JETP \textbf{27}, 793 (1968).

\bibitem{Volovik70}
G. E. Volovik, Calculation of the critical current  for particles decaying into nonparallel excitations, Soviet Phys. JETP \textbf{31}, 1106 (1970).

\bibitem{Brooks77}
J. S. Brooks and R. J. Donnelly, The Calculated Thermodynamic Properties of Superfluid Helium-4, J. Phys. Chem. Ref. Data \textbf{6}, 51 (1977).

\bibitem{Rayfield68}
G. W. Rayfield, Study of the Ion-Vortex-Ring Transition, Phys. Rev. \textbf{168}, 222 (1968).

\bibitem[Muller and Krstulovic(2020)]{Muller20}
 N. P. M\"uller and G. Krstulovic, Kolmogorov and Kelvin wave cascades in a generalized model for quantum turbulence, Phys. Rev. B {\bf 102}, 134513 (2020).

\bibitem[Muller and Krstulovic(2022)]{Muller22}
 N. P. Müller and G. Krstulovic, Critical velocity for vortex nucleation and roton emission in a generalized model for superfluids,   Phys. Rev. B {\bf 105}, 014515 (2022).

\bibitem{Nozieres2004}
P. Nozi\`{e}res, Is the Roton in Superfluid $^4$He the Ghost of a Bragg Spot? 
J. Low Temp. Phys. \textbf{137}, 45 (2004).

\bibitem{Balibar2007}
S. Balibar, The Discovery of Superfluidity, J. Low Temp. Phys. \textbf{146}, 441 (2007).

\bibitem{PiaevskiiPlateau}
L. P. Pitaevski\u{\i}, Properties of the spectrum of elementary excitations near the disintegration threshold of the excitations,
Sov. Phys. JETP, \textbf{36}, 830 (1959). 

\bibitem{MarchenkoParshin}
V. I. Marchenko and A. Ya. Parshin, On the Energy Spectrum of Helium II, JETP Letters \textbf{87}, 327 (2008)

\bibitem[Awschalom and Schwarz(1984)]{Awschalom84}
D. D. Awschalom and K. W. Schwarz, Observation of a Remanent Vortex-Line Density in Superfluid Helium,  Phys. Rev. Lett. \textbf{52}, 49 (1984).

\bibitem[Zurek(1985)]{Zurek}
W. H. Zurek, Cosmological experiments in superfluid helium?
Nature \textbf{317}, 505 (1985).

\bibitem[Vinen and Skrbek(2014)]{Vinen14}
W. F. Vinen and L. Skrbek, Quantum turbulence generated by oscillating structures, Proc. Natl. Acad.
Sci. (USA) \textbf{111} (Suppl. 1), 4699 (2014).

\bibitem{Goto}
R. Goto, S. Fujiyama, H. Yano, Y. Nago, N. Hashimoto, K. Obara, O. Ishikawa, M. Tsubota, and T. Hata, Turbulence in Boundary Flow of
Superfluid $^4$He Triggered by Free Vortex Rings, 
Phys. Rev. Lett. \textbf{100}, 045301 (2008).

\bibitem{Maris}
H. J. Maris, Electrons in Liquid Helium, J. Phys. Soc. Japan \textbf{77}, 111008 (2008).

\bibitem{SpringettDonnelly}
B. E. Springett and R. J. Donnelly,  Pressure Dependence of the Radius of the Negative Ion in Helium II,   Phys. Rev. Lett. \textbf{17}, 364 (1966).

\bibitem{Ostermeier}
R. M. Ostermeier, Pressure Dependence of Charge Carrier Mobilities in Superfluid Helium, Phys. Rev. A \textbf{8} 514 (1973).

\bibitem[McClintock(1978)]{McClintock78}
P. V. E. McClintock, An apparatus for preparing isotopically pure He$^4$, Cryogenics \textbf{18}, 201 (1978).

\bibitem[Hendry and McClintock(1987)]{Hendry87}
P. C. Hendry and P. V. E. McClintock,  Continuous flow apparatus for preparing isotopically pure $^4$He, Cryogenics \textbf{27}, 131 (1987).

\bibitem{normal}
Such a drag force can be interpreted as a classical viscous force exerted by the normal component of $^4$He.

\bibitem{charged_rings}
A charged vortex ring is a structure consisting of a quantized vortex rings with an ion bubble trapped on the vortex core.

\bibitem[Berloff&RobertsVI(1999)]{BerloffRobertsVI-1999}
N. G. Berloff and P. H. Roberts, Motions in a bose condensate: VI. Vortices in a nonlocal model,  J. Phys. A: Math. Gen. {\bf 32}, 5611 (1999).

\bibitem{Huepe97}
C. Huepe and M.-\'E. Brachet, Solutions de nucl\'eation tourbillonnaires dans un modele d’\'ecoulement superfluide, C. R. Acad. Sci., S\'er. II B \textbf{325}, 195
(1997).

\bibitem{Rica2001}
S. Rica, A remark on the critical speed for vortex nucleation in the nonlinear Schr\"odinger equation, Physica D \textbf{148}, 221 (2001).

\bibitem{Winiecki99}
T. Winiecki, J. F. McCann, and C. S. Adams, Vortex structures in dilute quantum fluids, Europhys. Lett. \textbf{48}, 475 (1999).

\bibitem{Winiecki2000}
T. Winiecki, B. Jackson, J. F. McCann, and C. S. Adams, Vortex shedding and drag in dilute Bose–Einstein condensates, J. Phys. B: At., Mol. Opt. Phys. \textbf{33},
4069 (2000).

\bibitem{Pham2005}
C.-T. Pham, C. Nore, and M.-E. Brachet,“Boundary layers and emitted excitations in nonlinear Schr\"odinger superflow past a disk, Physica D \textbf{210}, 203
(2005).

\bibitem{Stagg16}
G. W. Stagg, R. W. Pattinson, C. F. Barenghi, and N. G. Parker, Critical velocity for vortex nucleation in a finite-temperature Bose gas, Phys. Rev. A \textbf{93},
023640 (2016).

\bibitem{Stagg17}
G. W. Stagg, N. G. Parker, and C. F. Barenghi, Superfluid boundary layer, Phys. Rev. Lett. \textbf{118}, 135301 (2017).

\bibitem{Frisch24}
T. Frisch, S. Nazarenko, and S. Rica, Superflow passing over a rough surface: Vortex nucleation, Phys. Rev. Fluids \textbf{9}, 024701 (2024).

\bibitem[Villois and Salman (2018)]{VilloisSalman}
A. Villois and H. Salman, Vortex nucleation limited mobility of free electron bubbles in the Gross-Pitaevskii model of a superfluid,  Phys. Rev. B \textbf{97}, 094507 (2018).

\bibitem{Gross63}
E. P. Gross, Hydrodynamics of a superfluid condensate, J. Math. Phys. \textbf{4}, 195 (1963).

\bibitem{Dupont-Roc90}
J. Dupont-Roc, M. Himbert, N. Pavloff, and J. Treiner, Inhomogeneous Liquid $^4$He: A Density Functional Approach with a Finite-Range interaction, J. Low Temp. Phys. \textbf{81}, 31 (1990).

\bibitem{Pomeau93}
Y. Pomeau and S. Rica, Model of Superflow with Rotons, Phys. Rev. Lett. \textbf{71}, 247 (1993).

\bibitem{Dalfovo95}
F. Dalfovo, A. Lastri, L. Pricaupenko, S. Stringari, and J. Treiner, Structural and dynamical properties of superfluid helium: A density-functional approach, Phys. Rev. B \textbf{52}, 1193 (1995).

\bibitem[Berloff(1999)]{Berloff1999}
N. G. Berloff, Nonlocal Nonlinear Schr\"odinger Equations as Models of Superfluidity, J. Low Temp. Phys. {\bf 116}, 359 (1999).

\bibitem[Berloff \textit{et al.}(2014)]{Berloff2014}
N. G. Berloff, M. Brachet, and N. P. Proukakis, Modeling quantum fluid dynamics at nonzero temperatures,  Proc. Natl. Acad. Sci. USA {\bf 111} (Suppl. 1), 4675 (2014).

\bibitem[Reneuve \textit{et al.}(2018)]{Reneuve2018}
J. Reneuve, J. Salort, and L. Chevillard, Structure, dynamics, and reconnection of vortices in a nonlocal model of superfluids,  Phys. Rev. Fluids {\bf 3}, 114602 (2018).

\bibitem{Lahaye2009}
T. Lahaye, C. Menotti, L. Santos, M. Lewenstein, T. Pfau, The physics of dipolar bosonic quantum gases. Rep. Prog. Phys. \textbf{72}, 126401 (2009).

\bibitem{Parker2008}
N. G. Parker, D. H. J. O’Dell, Thomas–Fermi versus one- and two-dimensional regimes of a trapped dipolar Bose–Einstein condensate. Phys. Rev. A \textbf{78}, 041601R (2008).

\bibitem{Villois2025}
A. Villois, M. Onorato, and D. Proment, Vortex to Roton Transition in Dipolar Bose-Einstein Condensates, Phys. Rev. Lett. \textbf{134}, 253401 (2025).

\bibitem[Tuckerman \textit{et al.} (2004)]{Tuckerman2004}
L. Tuckerman, C. Huepe, and M.-E. Brachet, in Instabilities and Nonequilibrium Structures IX (Kluwer, Dordrecht, 2004), pp. 75–83.

\bibitem[Donnelly (1991)]{Donnelly1991}
R. J. Donnelly, \textit{Quantized Vortices in Helium II} (Cambridge University Press, 1991).

\bibitem{Krstulovic2020}
G. Krstulovic, \textit{A theoretical description of vortex dynamics in superfluids. Kelvin waves, reconnections and particle-vortex interaction}, Habilitation à diriger des recherches,
(Université Côte d’Azur, 2020), \url{https://hal.science/tel-03544830v1}.

\end{thebibliography}
\end{document}